\documentclass[aps,prapplied,twocolumn,superscriptaddress,amsmath,amssymb,preprintnumbers]{revtex4-2}

\usepackage{hyperref}
\usepackage{graphicx}
\usepackage{upgreek}

\newcommand{\micro}[0]{\(\upmu\)}
\newcommand{\ket}[1]{\ensuremath{\left| #1 \right\rangle}}
\newcommand{\vect}[1]{\boldsymbol{#1}}

\def\du{\ensuremath{\mathrm{d}}}
\def\eu{\ensuremath{\mathrm{e}}}
\def\iu{\ensuremath{\mathrm{i}}}

\hypersetup{
	colorlinks,
	citecolor=black,
	filecolor=black,
	linkcolor=black,
	urlcolor=black
}

\begin{document}

\title{Imaging current paths in silicon photovoltaic devices with a quantum diamond microscope}

\author{S.~C.~Scholten}
\affiliation{School of Physics, University of Melbourne, VIC 3010, Australia}

\author{G.~J.~Abrahams}
\affiliation{School of Physics, University of Melbourne, VIC 3010, Australia}

\author{B.~C.~Johnson}
\affiliation{School of Physics, University of Melbourne, VIC 3010, Australia}
\affiliation{School of Engineering, RMIT University, Melbourne VIC 3001, Australia}

\author{A.~J.~Healey}
\affiliation{School of Physics, University of Melbourne, VIC 3010, Australia}
\affiliation{Centre for Quantum Computation and Communication Technology, School of Physics, University of Melbourne, VIC 3010, Australia}

\author{I.~O.~Robertson}
\affiliation{School of Physics, University of Melbourne, VIC 3010, Australia}
\affiliation{School of Science, RMIT University, Melbourne VIC 3001, Australia}

\author{D.~A.~Simpson}
\affiliation{School of Physics, University of Melbourne, VIC 3010, Australia}

\author{A.~Stacey}
\affiliation{School of Physics, University of Melbourne, VIC 3010, Australia}
\affiliation{School of Science, RMIT University, Melbourne VIC 3001, Australia}

\author{S.~Onoda}
\affiliation{National Institutes for Quantum Science and Technology (QST), 1233 Watanuki, Takasaki, Gunma 370-1292, Japan}

\author{T.~Ohshima}
\affiliation{National Institutes for Quantum Science and Technology (QST), 1233 Watanuki, Takasaki, Gunma 370-1292, Japan}

\author{T.~C.~Kho}
\affiliation{School of Engineering, The Australian National University, Canberra, ACT 2601, Australia}

\author{J.~Ibarra~Michel}
\affiliation{Department of Electrical and Electronic Engineering, University of Melbourne, VIC 3010, Australia}

\author{J.~Bullock}
\affiliation{Department of Electrical and Electronic Engineering, University of Melbourne, VIC 3010, Australia}

\author{L.~C.~L.~Hollenberg}
\affiliation{School of Physics, University of Melbourne, VIC 3010, Australia}
\affiliation{Centre for Quantum Computation and Communication Technology, School of Physics, University of Melbourne, VIC 3010, Australia}

\author{J.-P.~Tetienne} 
\email{jean-philippe.tetienne@rmit.edu.au}
\affiliation{School of Physics, University of Melbourne, VIC 3010, Australia}	
\affiliation{Centre for Quantum Computation and Communication Technology, School of Physics, University of Melbourne, VIC 3010, Australia}
\affiliation{School of Science, RMIT University, Melbourne VIC 3001, Australia}

\date{\today}

\begin{abstract}
Magnetic imaging with nitrogen-vacancy centers in diamond, also known as quantum diamond microscopy, has emerged as a useful technique for the spatial mapping of charge currents in solid-state devices.
In this work, we investigate an application to photovoltaic (PV) devices, where the currents are induced by light.
We develop a widefield nitrogen-vacancy microscope that allows independent stimulus and measurement of the PV device, and test our system on a range of prototype crystalline silicon PV devices.
We first demonstrate micrometer-scale vector magnetic field imaging of custom PV devices illuminated by a focused laser spot, revealing the internal current paths in both short-circuit and open-circuit conditions.
We then demonstrate time-resolved imaging of photocurrents in an interdigitated back-contact solar cell, detecting current build-up and subsequent decay near the illumination point with microsecond resolution.
This work presents a versatile and accessible analysis platform that may find distinct application in research on emerging PV technologies.
\end{abstract}

\maketitle

\section{Introduction}\label{sec:intro}

Continued progress in photovoltaics technology critically relies on the development of new techniques to characterize photovoltaic (PV) materials and devices~\cite{wilson2020PhotovoltaicTechnologies2020}. 
In particular defects, carrier recombination sites and shunt resistances limit the power conversion efficiency of PV modules~\cite{breitensteinShuntTypesCrystalline2004,gauryChargedGrainBoundaries2017}, requiring spatially resolved investigation methods. 
To this end, a widely used technique is light beam induced current mapping  (LBIC), which measures the total photocurrent output by a device as a function of the position of a focused laser spot~\cite{padillaShortcircuitCurrentDensity2014,zookTheoryBeamInduced1983,michlImagingTechniquesQuantitative2014}.
While invaluable as a non-destructive tool to characterize efficiency variations across a PV module, LBIC necessitates a fully contacted device, and quantitative analysis of spatially localized carrier recombination events can be difficult as the technique measures a (spatially integrated) net current. Thermographic and luminescent~\cite{breitensteinLocalEfficiencyAnalysis2012} techniques are alternatively employed to detect defects~\cite{breitensteinDetectionShuntsSilicon2008} through mapping of locally dissipated power and the internal voltage (quasi-Fermi-level splitting) respectively. 
These techniques do not always require contacts, but as with LBIC extracting quantitative spatially resolved information is a challenge.

A potentially more direct way to gain insight into localized sources of inefficiencies or to measure device properties is by directly resolving the spatial distribution of charge flow (current) within the device. Such a mapping technique can be achieved, at least partially, using magnetic current imaging (MCI)~\cite{rockyPhotovoltaicModuleFault2019,kunzInvestigatingMetalsemiconductorContacts2019}. 
MCI measures the stray field in a parallel plane to the sample, linked to the current density through the Biot-Savart law, and thus can be performed without contacting the device.
In PV applications, the magnetic field map is measured using a scanning sensor such as a magnetic tunnel junction~\cite{schragCurrentDensityMapping2004,kogelMagneticFieldCurrent2016} or a Hall probe~\cite{rockyPhotovoltaicModuleFault2019}.
The sensor is often millimeter-sized, enabling imaging of whole cells or modules for fault detection~\cite{rockyPhotovoltaicModuleFault2019,kunzInvestigatingMetalsemiconductorContacts2019}.
Smaller sensors can map magnetic fields with a sub-micrometer resolution~\cite{schragScanningMagnetoresistiveMicroscopy2003}, but at the cost of mechanical delicacy and reduced sensitivity~\cite{scholtenWidefieldQuantumMicroscopy2021}. 
These sensors also cause optical shading, which has hindered their adoption for detailed quantitative analysis of PV devices. 

In this work, we employ a recently developed technology, quantum diamond microscopy (QDM), as an alternative method for realizing MCI of PV devices.
QDM exploits an array of optically addressable quantum sensors (nitrogen-vacancy centers) in diamond to generate a magnetic field image~\cite{levinePrinciplesTechniquesQuantum2019,rondinMagnetometryNitrogenvacancyDefects2014,scholtenWidefieldQuantumMicroscopy2021}. 
Compared with other magnetic imaging techniques, QDM presents several advantages appealing for PV applications.
QDM allows high-sensitivity calibration-free vector magnetic field mapping, ideal for accurate analysis of current distributions.
It offers diffraction limited spatial resolution (down to 500\,nm) over a wide field of view (up to several millimeters) in a robust imaging platform (no scanning) with good versatility to different operating conditions~\cite{scholtenWidefieldQuantumMicroscopy2021}.
Importantly, the diamond sensor is transparent, allowing illumination on the same surface of the cell as the mapping, without shading. 
QDM therefore appears promising as a tool for minimally-invasive quantitative characterization of PV devices at the micrometer scale, for instance to aid the analysis of recombination processes at defects or near contacts.

QDM has previously been applied to MCI in the context of (non-photoactive) solid-state devices, including small-scale (microns to tens of microns in size) graphene~\cite{tetienneQuantumImagingCurrent2017,lillieImagingGrapheneFieldeffect2019,kuImagingViscousFlow2020} and superconducting devices~\cite{lillieLaserModulationSuperconductivity2020} as well as larger (order millimeter) integrated circuits~\cite{nowodzinskiNitrogenvacancyCentersDiamond2015,turnerMagneticFieldFingerprinting2020,kehayiasMeasurementSimulationMagnetic2022}. 
Additionally, a recent investigation has used QDM for spatiotemporal mapping of photothermal vortices in MoS\(_2\)~\cite{zhouSpatiotemporalMappingPhotocurrent2020}.
However, so far the technique has not been used to map the photoresponse of a PV device. 
Here we use QDM to image magnetic fields induced by local laser illumination in crystalline silicon PV devices.
We first describe the experimental design and measurements protocols employed, before presenting results under both short-circuit and open-circuit conditions, illustrating how currents can be detected including closed loops entirely within the device.
We then demonstrate time-resolved imaging, allowing us to track the evolution of the current distribution upon illumination by a laser pulse.
We finish with a discussion of the challenges and opportunities for further application of QDM to PV research.

\section{Experimental methods}\label{sec:exp_design}

\begin{figure}[tb]
	\includegraphics[width=0.45\textwidth]{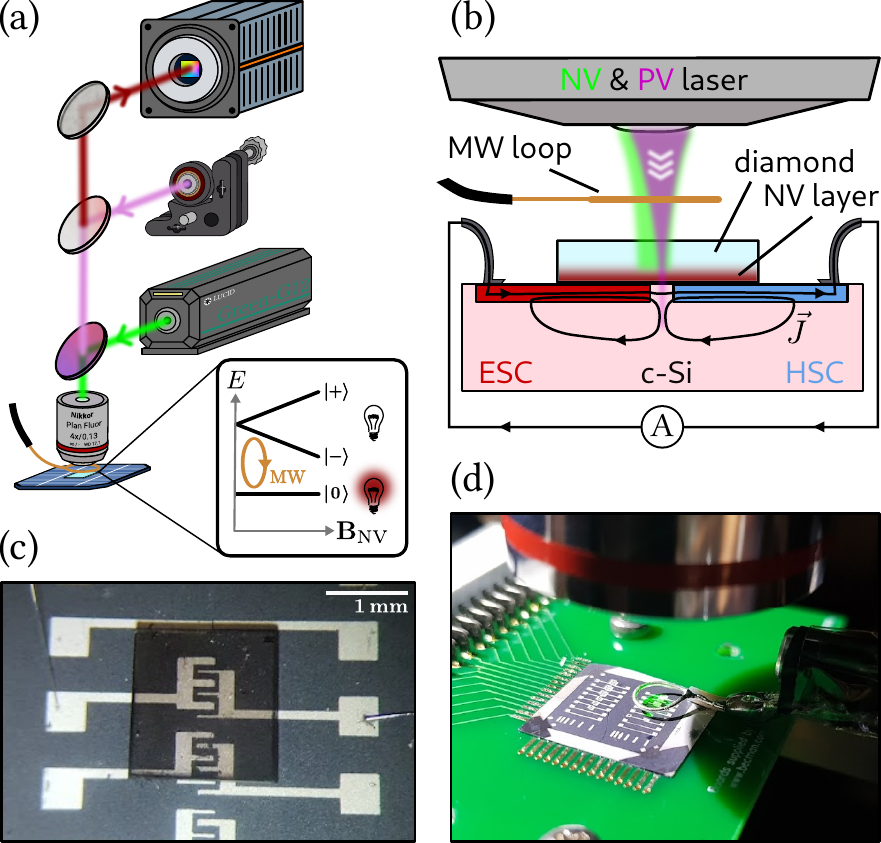}
	\caption{
		Imaging PV cells with a quantum diamond microscope.
		(a)~Schematic of experiment.
		A diamond containing a thin NV layer is placed on top of the PV cell. The PV cell is stimulated with an infrared laser (pink beam), while the NVs are controlled with a green laser (green beam).
		The NVs are readout via their photoluminescence (red beam) onto a camera.
		Magnetic imaging is achieved through measurement of the NV's magnetic resonance in a process termed optically detected magnetic resonance (ODMR; inset).
		(b)~Measurement geometry. 
		The PV laser excites a current in the device (black arrows), which produces a magnetic field that is measured by the proximal NV layer.
		Excited charges are collected selectively at the hole/electron selective contacts (H/ESC).
		(c)~Photograph of diamond sensing chip on top of custom silicon PV device. 
		(d)~Wider view of sensing geometry, showing the microwave (MW) loop.}\label{fig:schematic}
\end{figure}

\subsection{Experimental set-up}

The principle of the experiment is depicted schematically in Fig.~\ref{fig:schematic}(a,b). 
A diamond containing a thin layer of nitrogen-vacancy (NV) centers is placed atop a solar cell.
Charge carriers are excited in the bulk of the cell with an infrared (IR) laser ($\lambda=854$\,nm, ``PV laser''), which are extracted by the hole-selective and electron-selective contacts (HSC and ESC respectively).
This selectivity instantaneously produces a dipole across each junction.
The separated carriers then follow all return paths to equilibrate this dipole, distributed according to each path's potential and resistance~\cite{redfernInterpretationCurrentFlow2006}.
One of these return paths is the circuit outside the device (i.e.\ the load; here called the ``external path''), though additional loops may form within the device itself (``internal paths'', or shunts) [Fig.~\ref{fig:schematic}(b)].
In the presence of this selectivity, the photoexcited charge carrier flow produces a net current, which in turn produces a magnetic field.
The goal of the experiment is to image these current-induced magnetic fields, utilizing proximal NV centers as vector magnetometers, and then to reconstruct the source current density.
The NVs are excited with a green laser (\(\lambda=532\)\,nm, ``NV laser''), and their spin state read out via their red photoluminescence onto a camera.
Quantitative magnetic imaging is achieved by sweeping the frequency of an applied microwave field to obtain an optically detected magnetic resonance (ODMR) spectrum~\cite{rondinMagnetometryNitrogenvacancyDefects2014}.
A bias magnetic field is applied such that its projection onto each of the four NV orientations in the diamond are distinct.
In this way, the full vector field can be calculated from the measured projection of the sample's stray field onto each NV orientation~\cite{maertzVectorMagneticField2010}.

The devices probed in this work are crystalline silicon (c-Si) solar cells in which both contacts are placed on the rear of the cell [top side in Fig.~\ref{fig:schematic}(b)].
This so-called back-contact architecture is commonly employed in commercial solar cells to maximize light coupling into the c-Si absorber under normal front illumination~\cite{allenPassivatingContactsCrystalline2019} [bottom side in Fig.~\ref{fig:schematic}(b)].
The back-contact design also serves as a convenient test system for our experiment as it results in significant lateral transport, whereas the simpler sandwich design (one contact on either side of the absorber) is dominated by vertical transport, which is more difficult to probe using external magnetic field measurements.
Here we will mainly use back illumination, which allows us to easily focus a laser spot at a known location in the plane of the contacts, thus providing a relatively simple scenario facilitating analysis in this proof of concept.
However, front illumination is also possible as we demonstrate in Appendix~\ref{app:front}, and in principle our method can accommodate any illumination distribution (including uniform over the full cell area).  

Figure~\ref{fig:schematic}(c) shows a photograph of custom PV devices used in our experiments (see fabrication details in Appendix~\ref{app:pv_devices}), featuring a simple contact geometry. 
The diamond chip placed on top is 2\,mm\,$\times$\,2\,mm in size, which sets the maximum field of view achievable.
The diamond/PV device assembly as mounted in the optical microscope is shown in Fig.~\ref{fig:schematic}(d), with the microwave (MW) loop antenna visible between the device and the microscope objective.

\subsection{Measurement protocol}\label{sec:meas_protocol}

\begin{figure*}[tb]
	\includegraphics[width=0.85\textwidth]{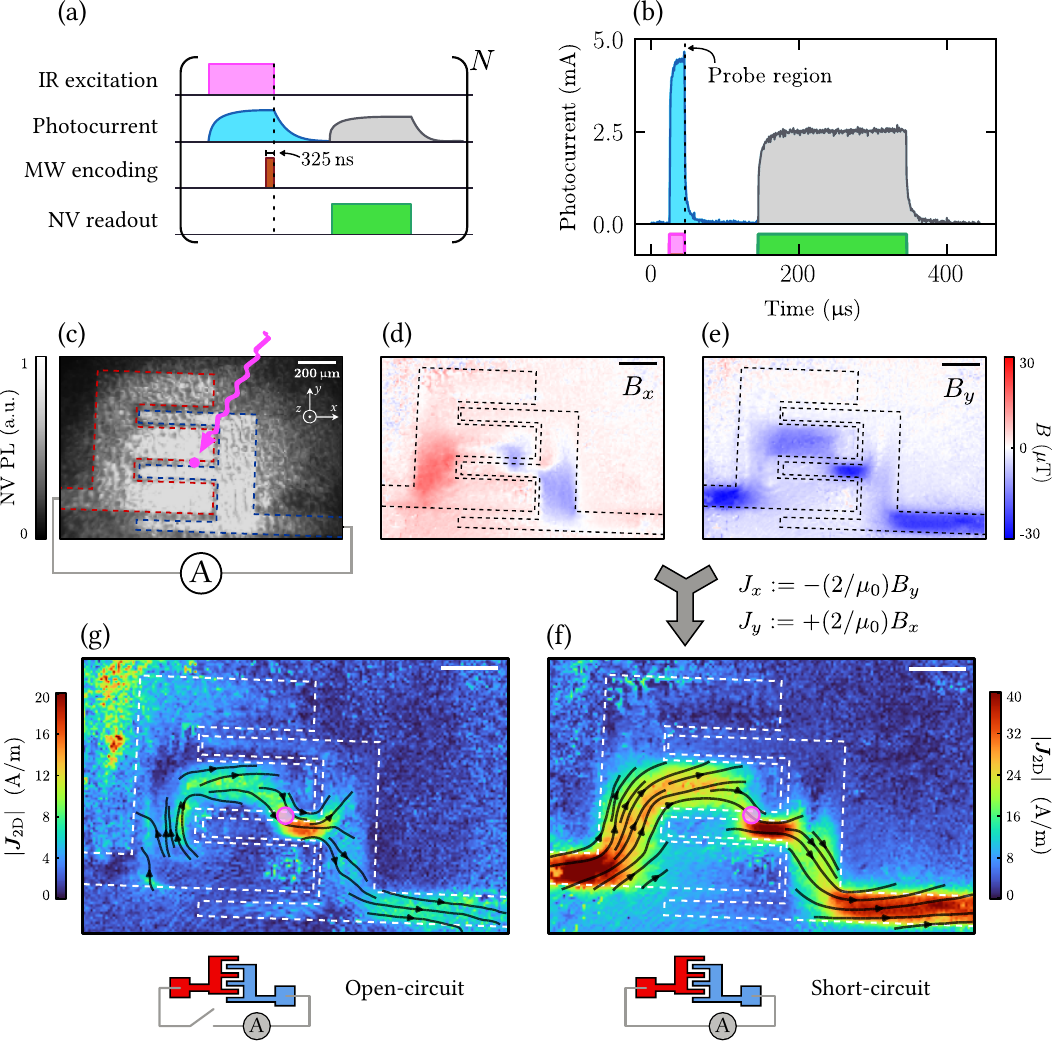}
	\caption{
		Photocurrent sensing protocol. 
		(a)~Stroboscopic pulse sequence used for magnetic photocurrent sensing.
		The sequence is only sensitive to magnetic fields during the MW pulse (duration 325\,ns), which is aligned with the back of the excitation pulse where the photocurrent is expected to be largest. 
		The green readout laser is temporally separated from the IR excitation laser to isolate the current produced by the later. 
		Note the IR laser has no effect on the NVs, and any IR reflections are filtered before reaching the camera.
		(b)~Electrically measured photocurrent through the pulse sequence, illustrating the temporal separation between pulses. 
		The region sensitive to magnetic fields is again annotated.
		(c)~Photoluminescence (PL) image of the sample photographed in Fig.\ref{fig:schematic}(c,d).
		Dashed lines indicate the extent of the contacts (red: ESC, blue: HSC). 
		The IR excitation location is illustrated in pink. 
		(d,e)~Vector magnetic field images taken in the same position as in (a), under short-circuit conditions. 
		These maps are calculated via imaging with an IR pulse as in (b), subtracted from an image taken without the IR pulse to remove contributions to the magnetic field not produced by the photocurrent.
		The coordinate system is indicated inset in (a). 
		Dashed lines indicate the extent of the contacts. 
		(f)~Current density map converted from the magnetic field images in (d,e) (see text for details). 
		Color indicates the magnitude of the current density, with the black streamlines providing vector information.
		A pink circle indicates the position of the IR excitation.
		(g)~Current density image taken in the same conditions as (f), except in open-circuit conditions.
		The apparent current in the top-left is an imaging artifact due to poor signal at the edge of the green illumination beam.
		Note the color scale has half the range of (f).}\label{fig:trident}
\end{figure*}

One challenge of applying QDM to interrogate a PV device is that the NV laser may induce additional photocurrents and therefore perturb the device under study.
One solution is to add a metal coating to the diamond so as to obstruct the NV laser~\cite{broadwayImagingDomainReversal2020}, but this laser shield would restrict illumination to the front face of the PV device.
Alternatively, it is possible to account for the effect of the NV laser by separating in time the PV excitation from the NV readout. 

The protocol we implement is depicted in Fig.~\ref{fig:trident}(a), which shows the pulse sequence (repeated continuously) and the expected time evolution of the photocurrent induced by both lasers. 
The PV device is first excited by an IR pulse during which a photocurrent builds up. 
A MW pulse is applied to probe the \(\ket{0} \leftrightarrow \ket{\pm}\) spin transitions of the NVs [Fig.~\ref{fig:schematic}(a); inset]. 
This MW pulse encodes the magnetic field information (contained in the spin transition frequencies via the Zeeman effect) into the NVs, which can be read out at a later time (limited by the NV spin lifetime, \(T_1\approx1\)\,ms). 
The green laser pulse subsequently reads out the NVs' spin state, exploiting the fact that the photoluminescence of the \(\ket{0}\) state is brighter than that of \(\ket{\pm}\)~\cite{rondinMagnetometryNitrogenvacancyDefects2014}, and re-initializes the NVs into the \(\ket{0}\) state. 
But this readout pulse also induces an additional (incidental) photocurrent.
A dark time is therefore inserted after the readout pulse and before the next IR pulse begins, to allow this photocurrent to fully decay and the device to return to equilibrium.

The photocurrent measured externally with an oscilloscope during the pulse sequence is shown in Fig.~\ref{fig:trident}(b) for the device of Fig.~\ref{fig:schematic}(c).
One can see that the photocurrent reaches a near-equilibrium at the end of the 20\,\(\upmu\)s IR pulse, and decays over a similar time scale.
The green pulse (200\,\(\upmu\)s) induces a current amplitude half the IR pulse and exhibits a similar rise/fall time.
Here we used a dark time of 100\,\(\upmu\)s between the green and IR pulses, which is sufficient to efficiently separate the two stimuli. 

We note that the MW encoding pulse is short (\(\approx 300\)\,ns) compared to the dynamics of the current, which allows us to perform stroboscopic imaging.
We initially place the MW pulse at the end of the IR pulse where the photocurrent is largest as shown in Fig.~\ref{fig:trident}(a,b), leaving a time-dependent study for Sec. \ref{sec:spatiotemporal_mapping}.
For each measurement, a reference image is acquired without the IR pulse, to normalize out any other field contributions. 
This process is undertaken in parallel for all pixels of the image and integrated over hours to improve the signal-to-noise ratio.

\section{Short-circuit current mapping}\label{sec:short_circ_mapping}

The photocurrent sensing protocol was first employed on a simple custom device with an interlocking trident geometry, shown in Fig.~\ref{fig:schematic}(c). It is comprised of an intrinsic c-Si bulk absorber, on which MoO\(_x\) (hole-selective), LiF (electron-selective), and aluminum (metallization) layers are evaporated.
The device is then contacted with aluminum wirebonds, and connected in short-circuit through a digital oscilloscope, which is used to calibrate the pulse sequence [Fig.~\ref{fig:trident}(b)].
A photoluminescence map of the NV layer above the device is shown in Fig.~\ref{fig:trident}(c), along with example vector magnetic field measurements (here $B_x$ and $B_y$ components are shown) in Fig.~\ref{fig:trident}(d,e).

The \O rsted magnetic field generated by the current 
\(\vect{J}(\vect{r})\) is described by the Biot-Savart law 
\begin{equation} \label{eq:biotsavart}
	\vect{B}(\vect{r}) = \frac{\upmu_0}{4\uppi} \int \du^3 \vect{r'}\  \frac{\vect{J}(\vect{r}') \times (\vect{r} - \vect{r}')}{|\vect{r} - \vect{r}'|^3},
\end{equation}
where \(\upmu_0\) is the vacuum permeability and the integral is over all space. 
In general, there is no unique solution for \(\vect{J}(\vect{r})\) when the current extends over a three-dimensional (3D) volume, as is the case here (the light penetrates some depth into the absorber, and the current paths extend throughout the entire thickness).
Nevertheless under appropriate assumptions it is possible to extract useful information using knowledge of the source or sample geometry~\cite{hamalainenMagnetoencephalographyTheoryInstrumentation1993}.
However, such 3D reconstruction methods are beyond the scope of this work. 
Instead, here we will make simplifying assumptions that allow us to represent the current density in a 2D form.

If the current is confined to a 2D plane, the current density is reduced to \(\vect{J}_{\rm 2D} = [J_x(x,y),J_y(x,y),0]\) and 
we can invert Eq.~\ref{eq:biotsavart} in Fourier space (taking the unitary, \(-i\) convention)~\cite{rothUsingMagnetometerImage1989,broadwayImprovedCurrentDensity2020}
\begin{equation}\label{eq:fourier_bxy2j}
	(B_x,~ B_y) \rightarrow
	\left\lbrace
		\begin{aligned}
			\mathcal{J}_x &=  -\alpha \mathcal{B}_y\\
			\mathcal{J}_y &=  +\alpha \mathcal{B}_x
		\end{aligned}\ .
		\right.
\end{equation}
Here \(\mathcal{J}(k_x, k_y)\) is the Fourier-space (2D) current density, \(\mathcal{B}(k_x, k_y)\) is the Fourier-space magnetic field at the NV plane, \(k_x\) and \(k_y\) are the in-plane angular wavenumbers with \(k = \sqrt{k_x^2 + k_y^2}\), and \(\alpha = 2 {\rm e}^{k \mathnormal{\Delta}} / \upmu_0\) for a sample-sensor offset distance of \( \mathnormal{\Delta}\).
This method utilizes the vector capability of our ensemble measurement to employ the in-plane field only, improving accuracy \cite{broadwayImprovedCurrentDensity2020}. 
Moreover, we can further simplify Eq.~\ref{eq:bxy2j} by taking the limit \(k \mathnormal{\Delta}\rightarrow 0\) which is valid here except for the highest spatial frequency of the image (of order the inverse of the pixel size, see Appendix \ref{app:curr_recon}). 
Making this simplification removes any amplification of high-frequency noise as well as any artefacts arising from Fourier transforms. 
Indeed, Eq.~\ref{eq:bxy2j} becomes a simple scaling of the magnetic field to current density units, in real space,
\begin{equation} \label{eq:bxy2j}
	\vect{J}_{\rm 2D} := \frac{2}{\upmu_0} (B_y, B_x, 0)~.
\end{equation}

Figure~\ref{fig:trident}(f) shows the resultant current density map, showing the expected pattern: conventional charge current flows from the ESC to the HSC, via the illumination point in the absorber.
The current flowing in the contacts is approximately conserved throughout the image, but appears reduced at the illumination point (little apparent current flowing between ESC and HSC). 
This apparent reduction is expected because the current there is delocalized in the $z$ direction into the absorber with a significant $J_z$ component, and is therefore not well captured by our simple model. 
In fact, the apparent lack of conservation of current in the $\vect{J}_{\rm 2D}$ map provides a signature of the 3D nature of the transport locally, and may form the basis for more sophisticated analysis methods for this scenario.  

\section{Open-circuit current mapping}\label{sec:open_circuit_mapping}

\begin{figure*}[tb]
	\includegraphics[width=0.95\textwidth]{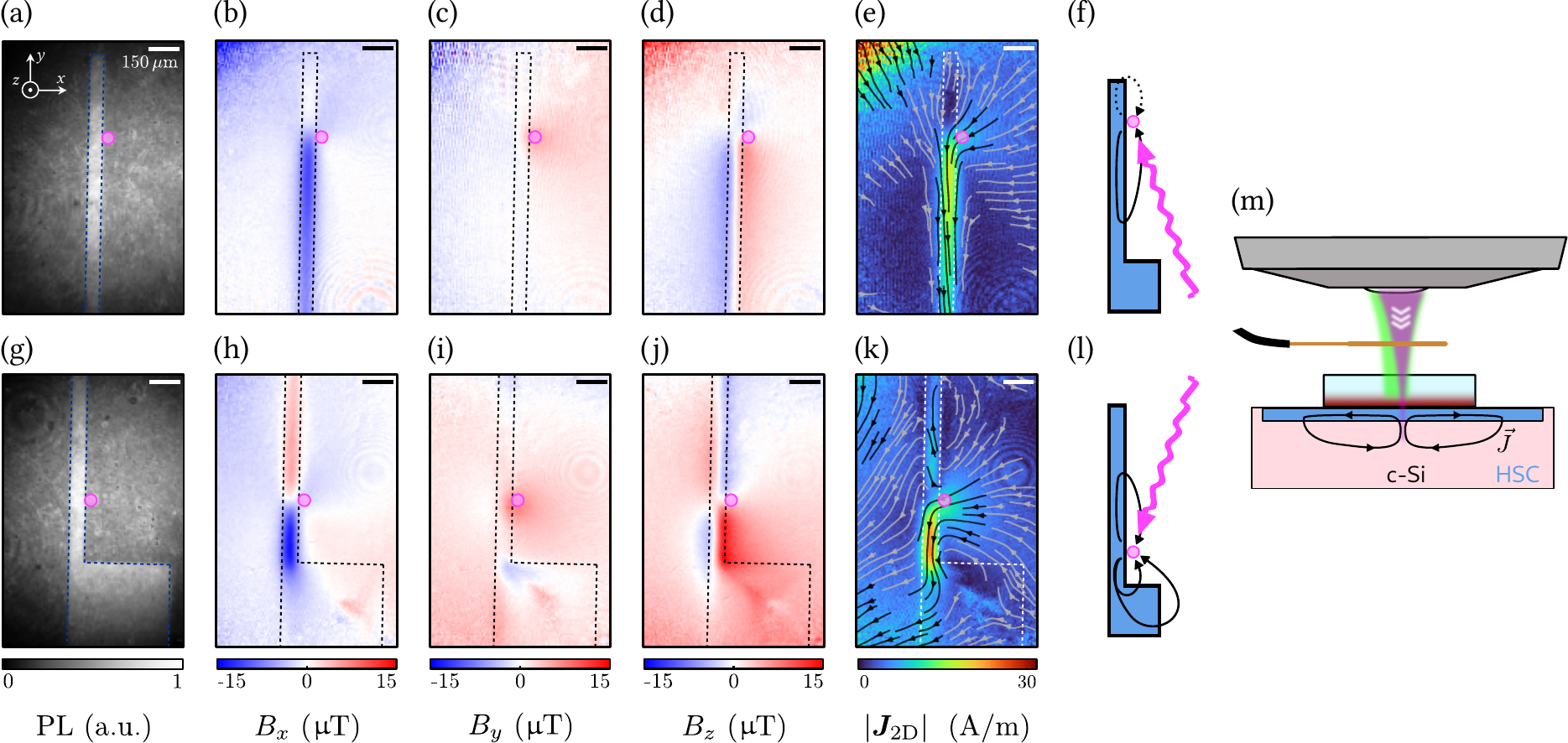}
	\caption{
		Imaging closed return currents in an isolated hole selective contact (HSC)\@.
		(a)~Photoluminescence image of the sample.
		The device has not been electrically contacted.
		The IR excitation location is illustrated in pink.
		(b-d)~Vector magnetic field images, taken under the same pulse sequence as in Fig.~\ref{fig:schematic}(b), normalized via an image taken without the IR pulse.
		(e)~Current density map calculated from (b-d), with black streamlines showing vector information for the large magnitude regions.
		Low current density magnitudes  are shown with gray streams to indicate the return path in the intrinsic region.
		(f)~Schematic of imaging geometry, indicating excitation location and theoretical return current paths.
		(g-l)~Images similar to (a-e), taken with IR excitation further down the contact, indicating the dependence of the current flow on the return path's resistance.
		(m)~Schematic of internal current loops for a single isolated contact.}\label{fig:solo}
\end{figure*}

We now investigate the presence of internal current paths under open-circuit conditions.
Figure~\ref{fig:trident}(g) is taken under identical conditions to Fig.~\ref{fig:trident}(f), except that the device is disconnected from any load.
The signal is significantly reduced and displays notably different qualitative behavior.
Note that no current is measured to be injected from the left of the image, i.e.\ the current is not flowing through the external circuit.
The picture is complicated by complex contact geometry, and the presence of both electron and hole selectivity whose local directionality add together.
This directionality can be seen at the excitation location, where conventional current flows from the ESC to the HSC and increases the distance of the return paths.

To simplify analysis and verify the existence of these internal current loops we now image isolated solo contacts.
In Fig.~\ref{fig:solo} we image an HSC under two different IR stimulus positions.
The top row [Figs.~\ref{fig:solo}(a-e)] is a measurement with stimulus near the top of the contact.
We present all components of the magnetic field in Figs.~\ref{fig:solo}(b-d) to show they are self-consistent: they indicate a strong current in the HSC in \(-\hat{y}\) away from the illumination point, and a dispersed current in the bulk [see Fig.~\ref{fig:solo}(e) for the reconstructed current density map].
We divide the possible return paths into two classes based on the sign of \(J_y\) in the contact, shown schematically in Fig.~\ref{fig:solo}(f).
The `positive' path current has a lower magnitude as the photogenerated excess carriers have less space (i.e. contact area) on this class of return route, and thus fewer paths and higher resistance~\cite{redfernInterpretationCurrentFlow2006}.
The current noticeably decays within the contact as it flows away from the stimulus, which in principle can be used to estimate the carrier (holes in this case) diffusion length.
A measurement is then taken with the illumination position moved down the contact towards the (unbonded) bonding pad, to illustrate a case of increased symmetry [Figs.~\ref{fig:solo}(h-j)].
The associated current map [Fig.~\ref{fig:solo}(k)] has a more pronounced `positive' path current, and the return flow through the bulk absorber is more evident.
The equivalent schematic to Fig.~\ref{fig:schematic}(b) can be drawn for this context, with two internal loops of opposite helicity within the one contact [Fig.~\ref{fig:solo}(m)]. 

These open-circuit experiments illustrate a useful aspect of MCI, which can detect internal current loops without the need for an external circuit. Importantly, these internal loops correspond to shunt resistances which, in normal solar cell operation, cause a loss of energy conversion efficiency. A tool allowing direct observation of these internal loops may therefore aid research into PV materials and cell designs.   

\section{Spatiotemporal current mapping}\label{sec:spatiotemporal_mapping}

\begin{figure*}[htb!]
	\includegraphics[width=0.86\textwidth]{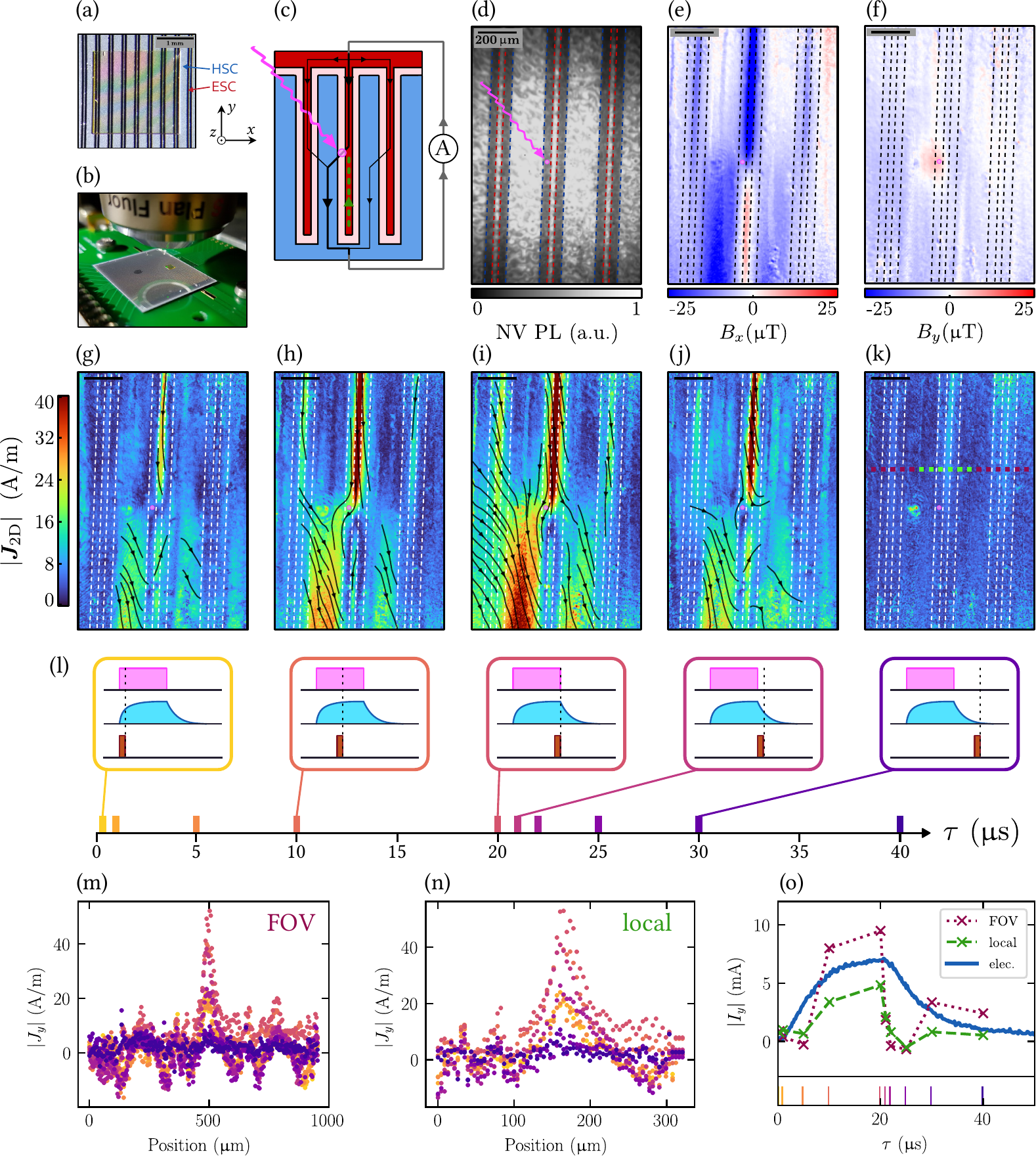}
	\caption{
		Spatiotemporal imaging of a lab-scale PV cell. 
		(a)~Image of diamond on PV cell. 
		The device is an interdigitated back contact (IBC) cell, illuminated (and imaged) from its back. 
		The hole-selective (blue; wide) and electron-selective (red; thin) contacts are labeled. 
		(b)~Wider view of diamond chip placed on cell. 
		(c)~Schematic of the cell with IR excitation location shown in pink. 
		Theoretical current paths under short-circuit conditions are annotated: black for the external path, green for an internal loop.
		(d)~Photoluminescence image of the device under test, annotated with the extent of the contacts via dashed lines, and the excitation region with a pink circle. 
		(e,f)~In-plane magnetic field component maps, taken at \(\tau = 20\)\,\micro s.
		(g-k)~Current density maps, each taken under the same conditions, except for the position of the MW pulse relative to the excitation pulse, each normalized by the equivalent pulse sequence before excitation. 
		As the MW pulse encodes the current-induced magnetic information, moving its position relative to excitation allows temporal imaging of current dynamics. 
		(l)~Timeline of images taken, from an origin \(\tau = 0\)\,\micro s equivalent to the MW pulse being immediately before the IR excitation pulse. 
		A schematic of the pulse sequence for each image in (g-k) is shown.
		(m)~Linecuts of \(J_y(y)\) across the full imaged field of view (\(\approx\)1\,mm), for the position indicated by the maroon dashed line in (k), for each image in (l). 
		(n)~Equivalent linecuts to (m), but shorter in extent, as indicated by the green dashed line in (k). 
		For both (m) and (n), the color of each scatter series indicates the image it was taken from, as indicated in (l). 
		(o)~Each series in (m)/(n) (maroon/green) is integrated to produce the net (\(y\)) current, as a function of time. 
		These series are plot against time, and compared to an electrical (full-device, external current only) measurement (blue line).
		The dashed lines below the plot identify each image, according to the scheme in (l).}\label{fig:tseries}
\end{figure*}

Advancing from test devices, we now analyze a larger-scale (\(2\times2\)\,cm\(^2\)) c-Si solar cell with high efficiency (24.4\% measured for a similar device~\cite{franklinDesignFabricationCharacterisation2016}) [Fig.~\ref{fig:tseries}(a,b)]. 
This device features a conventional interdigitated back contact geometry (see Appendix~\ref{app:pv_devices}), alternating ESC and HSC contact fingers with 50/180\,{\micro}m width.
Figure~\ref{fig:tseries}(c) schematically depicts the predicted current path: an internal loop (green dashed) at the end region of the closest ESC contact to the illumination (charges flowing anti-parallel to the external loop), and a broader distribution as the charges are collected over a larger surface area.
A PL image of the device and the illumination position is shown in Fig.~\ref{fig:tseries}(d); red (blue) dashed lines outline the extremities of the electron (hole) selective contact, which are all connected at the top (bottom) (\(\pm y\)) via a large busbar.
Initial field maps [Fig.~\ref{fig:tseries}(e,f)] follow the hypothesized pattern; the red region at the bottom of Fig.~\ref{fig:tseries}(e) represents current flow up in the dead end ESC finger, while the larger downward flow (blue) is from the central ESC to the left HSC, across the point of illumination.

As previously mentioned, the strobosocopic nature of the measurement protocol allows for temporal information to be extracted by changing the position of the MW pulse relative to photoexcitation.
To illustrate this capability, we display images taken at different times [Fig.~\ref{fig:tseries}(g-k)], as depicted schematically below each image in Fig.~\ref{fig:tseries}(l).
Here we use a measurement taken immediately before the IR pulse as a reference for all images in the series, unlike the previous figures that required one reference measurement per image.
Figures~\ref{fig:tseries}(g-k) thus represent the difference in current density from what was present immediately before the IR pulse, including any effects with longer timescales.
Overall the images follow a trend that appears largely simple: the current density builds up during the IR pulse and subsequently decays, conventional current flows from the ESC to the HSC predominantly through the excitation spot and appears roughly conserved (spatially) throughout.

However, on closer examination, more subtle aspects of the device's dynamics are revealed; for instance the current in the dead end ESC finger is seen to peak after the end of the IR pulse [compare Fig.~\ref{fig:tseries}(j) with Fig.~\ref{fig:tseries}(i)].  
To exemplify how spatially-dependent dynamics can be quantified, we take horizontal linecuts [annotated in Fig.~\ref{fig:tseries}(k)] 200\,\(\upmu\)m above the illumination point, the results for each image are shown as separate series in Figs.~\ref{fig:tseries}(m,n).
The full field of view [FOV; maroon linecut in Fig.~\ref{fig:tseries}(m)] data shows a large central feature corresponding to the ESC contact, with smaller signals in the other contacts.
There is a noticeable pattern of negative features, indicative of return currents.
Focusing on the large feature at the ESC [green linecut, Fig.~\ref{fig:tseries}(n)] identifies the current transience.
Numerically integrating these linecuts, and plotting as a function of time [Fig.~\ref{fig:tseries}(o)] allows us to compare the imaged current with that measured externally by a digital oscilloscope.
Both the FOV and local curves show roughly the same overall shape as the electrical measurement, but with some noticeable differences. In particular, the MCI-deduced current rises more sharply and after a delay of 5-10\,\(\upmu\)s compared to the electrically measured current, which can be understood as the latter includes carriers traveling some distance in the absorber, only reaching the contacts north of the linecuts position. Furthermore, the non-monotonic decay after the end of the IR pulse, as well as the magnitude of the FOV measurement (exceeding the electrically measured current) are the result of complex dynamics of internal return paths.
While a full analysis is beyond the scope of this work, these measurements illustrate the benefit of MCI as a quantitative tool to study the spatiotemporal dynamics of charge transport in PV devices.

\section{Discussion}\label{sec:discussion}

In this report we have demonstrated magnetic current imaging of crystalline silicon photovoltaic cells with a quantum diamond microscope.
Compared to other methods to realize MCI, this platform offers generally superior field sensitivity for imaging at micrometer scales~\cite{scholtenWidefieldQuantumMicroscopy2021}, is relatively simple to implement and robust to operate given the absence of moving parts, and is not subject to optical shading.
The accessibility of the technique can be further improved by automating the positioning of the diamond sensor near the device's surface~\cite{abrahamsIntegratedWidefieldProbe2021}, alleviating the need for the user to handle the diamond directly.
The size of typical diamond substrates and laser intensity requirements limits the field of view to a few millimeters at most~\cite{turnerMagneticFieldFingerprinting2020}, although imaging over centimeters is in principle achievable via image stitching.
However, as the carrier diffusion length in PV semiconductors rarely exceeds a few hundreds of micrometers (e.g. in c-Si solar cells), the QDM platform appears well suited for fundamental studies of PV devices under localized excitation, in which case all the relevant physics can be captured in a single image. 

For c-Si PV cells as studied in this work, where the current is dispersed throughout a thick absorber (100s of \(\upmu\)m, commensurate with the carrier diffusion length), analysis of the magnetic field data remains a challenge as there is no unique solution for the current density.
Nevertheless, even a simple reconstruction algorithm which assumes a planar current source, as employed here, is successful at revealing key features of the device's behavior, such as the internal closed current paths associated with shunt resistances observed in our experiments. This type of analysis may be useful, for example, to investigate shunt pathways in cell architectures with overlapping contact regions~\cite{tomasiSimpleProcessingBackcontacted2017}. 
There is also ongoing work to improve current reconstruction methods in MCI.
Clement et al.~\cite{clementReconstructionCurrentDensities2019} have developed a Bayesian technique to model currents at the image boundary, which are otherwise a major source of artifacts. 
Garsi et al.~\cite{garsiNoninvasiveImagingThreedimensional2021} have tackled the 3D reconstruction problem for the analysis of integrated circuits. 
Their method groups current sources (modeled as elements of an infinite wire) into different depth classifications, and then runs the 2D reconstruction algorithm on each plane of sources separately.
Such a technique could potentially be extended to the analysis of depth currents in a PV cell, although the lack of stratified structure (unlike an integrated circuit) poses a serious challenge. Using detailed models of the charge transport in the device to extract useful information, inspired by methods used in magnetoencephalography  ~\cite{hamalainenMagnetoencephalographyTheoryInstrumentation1993}, could be another interesting avenue. Here QDM has an advantage over other MCI implementations, as previous work has shown that vector magnetometry provides superior information for reconstructing source fields~\cite{broadwayImprovedCurrentDensity2020}.

In the PV research space, a promising application of QDM is as a precision analysis tool for the imaging of thin-film devices, such as perovskite solar cells. 
Indeed, for such devices the absorber layer is typically less than a micrometer in thickness, such that 2D reconstruction algorithms may be employed while introducing little error.
Perovskites bring additional technical challenges for QDM, such as overlap between the NV and perovskite absorption spectra, but the latter can be circumvented by temporal (via an appropriate pulse sequence) or spatial separation (by shielding the NV laser and using front illumination). 
However, thin-film PV devices generally feature much shorter diffusion lengths (\(\approx\)\,1\,\(\upmu\)m) compared to c-Si, as well as geometrically complex electrode structures~\cite{houBackcontactPerovskiteSolar2018,huPerovskiteSolarCells2019}, and as such will require the spatial resolution to be optimized. 
QDM experiments with this goal would need to prioritize achieving at least diffraction limited resolution ($\approx500$\,nm), requiring careful reduction of standoffs~\cite{abrahamsIntegratedWidefieldProbe2021} or perhaps super-resolution techniques~\cite{chenSuperresolutionMultifunctionalSensing2019}. 
Finally, while interdigitated back-contact devices are an obvious target with their predominantly lateral transport, sandwich-type devices with contacts on either side of the perovskite film could also serve as an interesting test system to explore current inversion methods adapted to this regime, where transport occurs primarily in the vertical direction ($J_z$; see Appendix~\ref{app:curr_recon}).

A few principal use cases of our technique present themselves.
First is the imaging of local excitation phenomena (such as the internal currents presented here) that are difficult to simulate, with an eye to improving theoretical models.
The second opportunity comprises measurement without optical excitation alongside application of a voltage across specific contacts, and then comparison of the image to a diode model for the extraction of device parameters~\cite{kunzInvestigatingMetalsemiconductorContacts2019}.
The latter application would not require current reconstruction methods, with theory and model compared via magnetic field alone. Beyond looking at complete PV devices, QDM could also be useful as a tool to study photo-induced transport at different stages of a device's fabrication. For example, the imaging of pinhole defects in passivating layers on c-Si~\cite{tetzlaffSimpleMethodPinhole2017}, a technologically important problem, may be possible without requiring any contact. In the same spirit, the results of Fig.~\ref{fig:solo} showed the value of measuring a device with a single contact type (here hole selective), and it should also be possible to extract materials parameters using only non-selective contacts (in which photocurrents may arise from geometrical asymmetries). More broadly, there is ample scope to apply QDM to study transport in a range of materials under localized optical excitation, from 2D materials to organic thin films.

\begin{acknowledgments}
This work was supported by the Australian Research Council (ARC) through grants CE170100012, DE190100336, DP190101506, FT200100073 and DP220100178.
S.C.S. gratefully acknowledges the support of an Ernst and Grace Matthaei scholarship.
A.J.H. is supported by an Australian Government Research Training Program Scholarship. J.B. acknowledges support from the Australian Centre for Advanced Photovoltaics Fellowship.
Part of this study was supported by QST President’s Strategic Grant ``QST International Research Initiative''.

\end{acknowledgments}

\bibliography{pv}	

\appendix

\section{Diamond samples}\label{app:diamond_samples}

The NV-diamond sample used in this work was made from a \(2\,{\rm mm}\times 2\,{\rm mm}\times 200\,\upmu{\rm m}\) electronic-grade chemical vapor deposition (CVD) diamond substrate, with [100]-oriented polished faces.
A 10\,\(\upmu\)m thick layer of N-doped isotopically enriched (99.95\% \(^{12}\)C) diamond was overgrown via microwave plasma-assisted CVD.
To increase the density of NV centers in the grown layer, the diamond was irradiated with 2\,MeV electrons (1\,\(\times\)\,10\(^{18}\) cm\(^{-2}\) fluence) and annealed using a ramp sequence culminating at 1100\,\(^{\circ}\)C to maximize NV yield and ensemble coherence properties~\cite{tetienneSpinPropertiesDense2018}. After annealing the chip was acid cleaned (\(30\)\,minutes in a boiling mixture of sulfuric acid and sodium nitrate).
The 10\,\(\upmu\)m thick sensing layer limits spatial resolution when imaging the current in the metal contacts as the standoff is comparatively small (\(\approx\,3\,\upmu\)m estimated between diamond surface and contacts), but has negligible influence where the current is dispersed in the bulk. This NV layer thickness was chosen as a tradeoff between sensitivity and spatial resolution~\cite{scholtenWidefieldQuantumMicroscopy2021}.

\section{PV devices}\label{app:pv_devices}

\begin{figure}[b!]
	\includegraphics[width=0.40\textwidth]{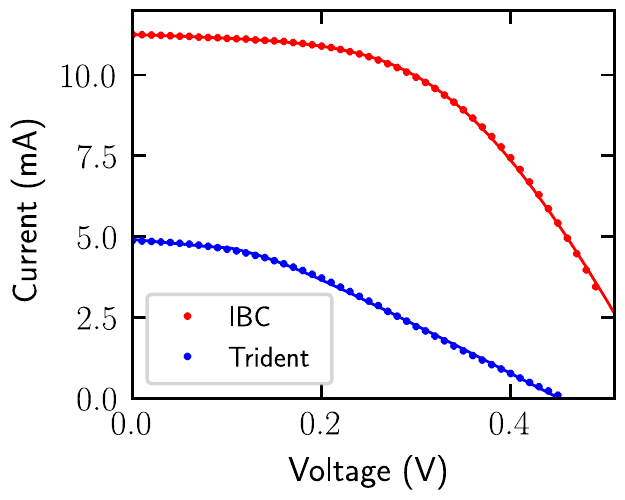}
	\caption{IV curves under local IR illumination for the trident cell imaged in Fig.~\ref{fig:trident} (blue) and the IBC cell imaged in Fig.~\ref{fig:tseries} (red).
	The lines are fits to a single diode model following \textcite{jainExactAnalyticalSolutions2004}.}\label{fig:app_iv}
\end{figure}

The devices imaged in Figs. 2 and 3 were fabricated with a $350\,\mu$m thick, 1-10\,\(\Omega\).cm n-type \(\langle100\rangle\) silicon wafer. The contacts were formed by evaporating MoO\(_x\) (10 nm) and LiF (3 nm) through a shadow mask to form a hole or an electron selective contact~\cite{bullockEfficientSiliconSolar2016}.
Evaporation was performed under a vacuum of \(1\times10^{-6}\)\,mbar and was immediately followed by a 200\,nm Al evaporation in both cases. 

The IBC cell imaged in Fig.~\ref{fig:tseries} was constructed from a $230\,\mu$m thick 1.5\,\(\Omega\).cm n-type Czochralski silicon wafer.
The ESCs are made up of localized diffuse POCl\(_3\) back-surface field regions and the HSCs formed from diffused BBr\(_3\) sheet emitters,  each passivated with SiO\(_2\)/Si\(_3\)N\(_4\) layers.
The aluminum metallization contacts locally through lithographic holes in the passivation. The front surface was passivated with SiN$_x$ and further coated with SiO$_x$ to form an anti-reflection coating. Further details can be found in Franklin et al.~\cite{franklinDesignFabricationCharacterisation2016}.

An exemplary IV curve for each device, under localized IR laser illumination, is shown in Fig.~\ref{fig:app_iv}.

\section{NV microscope setup}\label{app:nv_mic_setup}

All experiments were conducted with a purpose-built widefield NV microscope, depicted schematically in Fig.~\ref{fig:schematic} of the main text.
Optical excitation and initialization of the spins was achieved with a continuous wave \(\lambda = 532\)\,nm laser (Coherent Verdi), gated using an acousto-optic modulator (AA Opto-Electronic MQ180-AO,25-VIS), and focused using a widefield lens (\(f = 200\)\,mm) onto the back aperture of a microscope objective (Nikon Plan Fluor \(4\times\), NA\,\(=\)\,0.3).
The total laser power at the sample was 200\,mW, with a spot diameter of 1\,mm.
The red PL from the NV defects was collected back through the same objective and separated from the excitation beam with a dichroic mirror, and filtered (see below) before being imaged using a tube lens (\(f = 300\)\,mm) onto a scientific CMOS camera (Andor Zyla 5.5-W USB3), giving a diffraction-limited resolution of \(\approx\)\,1.2\,\(\upmu\)m across a maximum FOV of 2\,mm.

The PV devices were stimulated with an 854\,nm laser (diode: Lumentum 22045504 350\,mW; controller: MOGLabs DLC202), cleaned with a filter, gated with an AOM (AA Opto-Electronic MT80-A1-IR), combined with the imaging (camera-side) path with a second dichroic and finally focused onto the imaging objective back aperture with a widefield lens (400\,mm).
The total power at the sample was 200\,mW, with a spot diameter of \(\approx\)\,50\,\(\upmu\)m.

The PV laser wavelength is chosen to avoid interaction with the NVs, both in absorption and photoluminescence.
First, the dichroic used to separate the NV excitation and photoluminescence beams is chosen to transmit the PV laser (longpass 600\,nm).
The second dichroic is chosen to transmit the NV PL and reflect the IR beam (shortpass 800\,nm).
The IR beam required cleaning (longpass 780\,nm) to remove spectral components within the NV absorption (\(450\)-\(550\)\,nm) and PL (\(600\)-\(850\)\,nm) bands.
Before the camera an additional filter is used to ensure imaging of only PL wavelengths (bandpass \(690 \pm 60\)\,nm).

Microwave (MW) driving was provided by a signal generator (Rohde \& Schwarz SMBV100A) gated using built-in IQ modulation and amplified (Amplifier Research 60S1G4A).
A pulse pattern generator (SpinCore PulseBlasterESR-PRO 500\,MHz) was used to gate the lasers (via their respective AOMs) and MW (via the IQ inputs), as well as trigger the camera acquisition.
The amplified output was connected to a coaxial cable terminator by a loop of thin copper wire, which was positioned closely above the sample.

A permanent magnet was used to apply a bias field \(\vect{B_0}\) aligned such that its projection onto each NV axis (the \(\langle111\rangle\) family) is sufficiently different to independently resolve each resonance.
The measurements reported here were taken under \(B_0 \approx 6\)\,mT.

\section{NV measurements}\label{app:nv_measurements}

Optically detected magnetic resonance spectra of the NV sensing layer were acquired by sweeping the MW frequency with a dwell time of 30\,ms per frequency, matched to the camera exposure time.
The ODMR spectrum is normalized by acquiring a camera frame with the MW off in between each MW frequency; the spectrum with MW on divided by that without.
Within each exposure the lasers and MW were continuously gated, forming many repetitions of the same pulse sequence per frame.
Each frequency sweep thus requires a few seconds (30\,ms times the number of frequency steps, e.g.\ 120, times two for normalization), which is then repeated thousands of times to improve the signal-to-noise ratio (SNR).
Total acquisition time for each measurement was at least six hours, doubled again if a reference measurement was required (e.g.\ without the PV laser).

Note that the AOM introduces an \(\approx\)\,200\,ns lag for the laser pulses with respect to the MW, with an additional 200\,ns rise/fall time.
The pulse timings in Fig.~4 of the main text match the timings at the pulse pattern generator and are not corrected for these short delays.

After acquisition the ODMR spectra (acquired simultaneously at each pixel on the camera) were fit with the sum of eight Lorentzian functions with free peak frequencies (\(f_i, i \in \{1,2, \dots 8\}\)), amplitudes and widths, using a least-squares fitting algorithm.
Extracting the fit parameters per-pixel allows for spatial maps of these quantities to be constructed.
The electronic ground state of the NV defect can be modeled by the following spin-1 Hamiltonian (in Hz)
\begin{equation}
    \mathcal{H} = D S_Z^2 + \gamma_{\rm NV} \vect{S} \cdot \vect{B}\,,\label{eq:hamil}
\end{equation}
where \(XYZ\) is the reference frame specific to each NV orientation (\(\langle 111 \rangle\); \(Z\) is the symmetry axis), \(\vect{S} = (S_X, S_Y, S_Z)\) are the spin-1 operators, \(D\approx2.870\)\,GHz is the temperature-dependent zero-field splitting, and \(\gamma_{\rm NV}=28.033(3)\)\,GHz/T is the gyromagnetic ratio of the defect.
The total magnetic field \(\vect{B}_{\rm tot}\) at the NVs is calculated by minimizing the RMS error between the fit frequencies \(f_i\) and those calculated theoretically from the Hamiltonian \(\mathcal{H}\).
Any background fields, for example that from the bias magnet, are normalized via a reference measurement (e.g.\ the same experiment without PV laser) \(\vect{B}_{\rm ref}\), which is then subtracted to give \(\vect{B} = \vect{B}_{\rm tot} - \vect{B}_{\rm ref}\).

\begin{figure*}[t!]
	\includegraphics[width=0.9\textwidth]{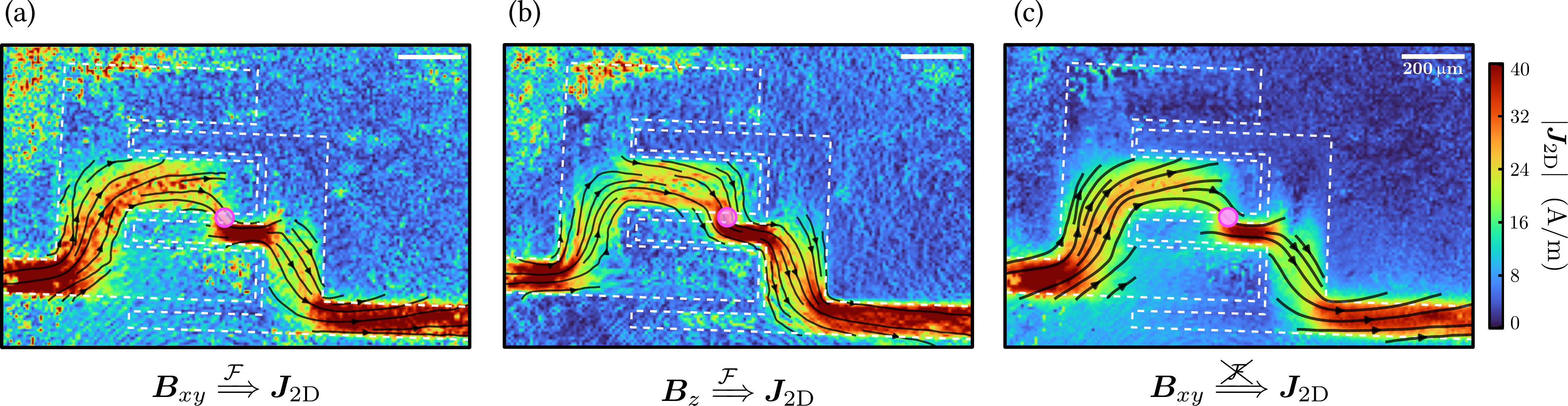}
    \caption{Comparison of current density reconstruction methods. The dataset in Fig.~\ref{fig:trident}(f) reconstructed from \(\vect{B}_{xy}\) [Eq.~\ref{eq:fourier_bxy2j}; (a)], from \(\vect{B}_{z}\) [Eq.~\ref{eq:fourier_bz2j}; (b)], and from \(\vect{B}_{xy}\) without a Fourier transform [Eq.~\ref{eq:bxy2j}; (c)].}\label{fig:apprecon}
\end{figure*}

\section{Current density reconstruction}\label{app:curr_recon}

The Biot-Savart law (Eq.~\ref{eq:biotsavart}) can be expressed in (2D) Fourier space as
\begin{align}
    \begin{split}
        \mathcal{B}_x(k_x, k_y, z) = 
        \frac{\upmu_0}{2} 
        \int_{-\infty}^{\infty} &
        \eu^{- k (z-z')}
        \Big [
        \mathcal{J}_y(k_x, k_y, z')\ +
        \\
        & \iu \frac{k_y}{k} \mathcal{J}_z(k_x, k_y, z') 
        \Big ] \du z'
    \end{split}\label{eq:fspace_biot_1} \\
    \begin{split}
        \mathcal{B}_y(k_x, k_y, z) = 
        \frac{\upmu_0}{2} 
        \int_{-\infty}^{\infty} &
        \eu^{- k (z-z')}
        \Big [
        - \mathcal{J}_x(k_x, k_y, z')\ -
        \\
        & \iu \frac{k_x}{k} \mathcal{J}_z(k_x, k_y, z') 
        \Big ] \du z'
    \end{split}\label{eq:fspace_biot_2} \\
    \begin{split}
        \mathcal{B}_z(k_x, k_y, z) = 
        \frac{\upmu_0}{2}
        \int_{-\infty}^{\infty} &
        \eu^{- k (z-z')}
        \Big [
        - \iu \frac{k_y}{k} \mathcal{J}_x(k_x, k_y, z')\ +
        \\
        & \iu \frac{k_x}{k} \mathcal{J}_y(k_x, k_y, z')
        \Big ] \du z'~.
    \end{split}\label{eq:fspace_biot_3}
\end{align}
Conventionally, the 2D source approximation is now made (or current confined within a slab with no \(z\) dependence), which leads to setting \(\mathcal{J}_z = 0\) everywhere.
The integrals in Eqs.~\ref{eq:fspace_biot_1}-\ref{eq:fspace_biot_3} can now be evaluated, and organized into a matrix equation (\(\mathnormal{\alpha}\) is defined in the main text):
\begin{equation} \label{eq:bs_matrix}
    \begin{bmatrix}
        \mathcal{B}_x \\ \mathcal{B}_y \\ \mathcal{B}_z
    \end{bmatrix}
    = 
    \frac{1}{\alpha}
    \begin{bmatrix}
        0& 1 \\
        -1& 0 \\
        -\iu k_y/k& \iu k_x/k
    \end{bmatrix}
    \begin{bmatrix}
        \mathcal{J}_x \\ \mathcal{J}_y
    \end{bmatrix}~.
\end{equation}
Equation~\ref{eq:bs_matrix} is over-constrained: there are several ways to deduce \(\vect{j}\).
The conspicuous choice is to reconstruct from \(B_z\) via
\begin{equation}\label{eq:fourier_bz2j}
	B_z \rightarrow
	\left\lbrace
		\begin{aligned}
			\mathcal{J}_x &=  \frac{\alpha k_y}{\iu k} \mathcal{B}_z\\
			\mathcal{J}_y &=  -\frac{\alpha k_x}{\iu k} \mathcal{B}_z
		\end{aligned}\,.
		\right.
\end{equation}
as Eq.~\ref{eq:fspace_biot_3} indicates it has no contributions from vertical currents (\(\mathcal{J}_z\)) that would contradict our 2D approximation.
However choosing to use \(B_z\) requires the use of current continuity (\(\nabla \cdot \vect{J} = 0\)), and then the assumption \(j_z = 0\) in order to eliminate either \(\mathcal{J}_x\) or \(\mathcal{J}_y\) and therefore it is impossible to avoid the influence of \(\mathcal{J}_z\) in any reconstruction choice.
Thus we choose to reconstruct from the in-plane magnetic field components, for the reasons following.
At this stage we also account for the thickness (\(t_{\rm NV}\)) of our sensing layer by averaging over source-sensor distances:
\begin{equation} \label{eq:thick_nv}
    (\mathcal{B}_x, \mathcal{B}_y) = \frac{\upmu_0}{2}\eu^{-k\mathnormal{\Delta}_{\rm NV}} \frac{\mathrm{sinh}(k t_{\rm NV} / 2)}{k t_{\rm NV} / 2} (\mathcal{J}_y, -\mathcal{J}_x)~,
\end{equation}
with \(\mathnormal{\Delta}_{\rm NV}\) the mean standoff distance.

We make the simplest additional approximation, taking \(k\mathnormal{\Delta}_{\rm NV} \rightarrow 0\), an equivalent statement to claiming the scale of the lateral variations in the image is much larger than the standoff.
In our experiment this is true, except at the smallest length scales.
We also take the similar approximation \(k t_{\rm NV} \rightarrow 0\).
The effect of these simplifications is an artificially (spatially) broadened signal as the highest spatial frequencies have not been adequately amplified.
The net result is Eq.~\ref{eq:bxy2j}, which requires no Fourier transform and thus avoids a separate set of processing artifacts~\cite{feldmannResolutionTwodimensionalCurrents2004, meltzerDirectReconstructionTwodimensional2017}. 
Note that reconstruction from \(B_z\) would require a Fourier transform, as well as introducing singularities in \(k\)-space~\cite{broadwayImprovedCurrentDensity2020}.
Figure~\ref{fig:apprecon} contains a comparison of reconstruction methods for the measurement in Fig.~\ref{fig:trident}(f).

\section{Isolating external currents}\label{app:isol_ext_current}

\begin{figure*}[t!]
	\includegraphics[width=0.9\textwidth]{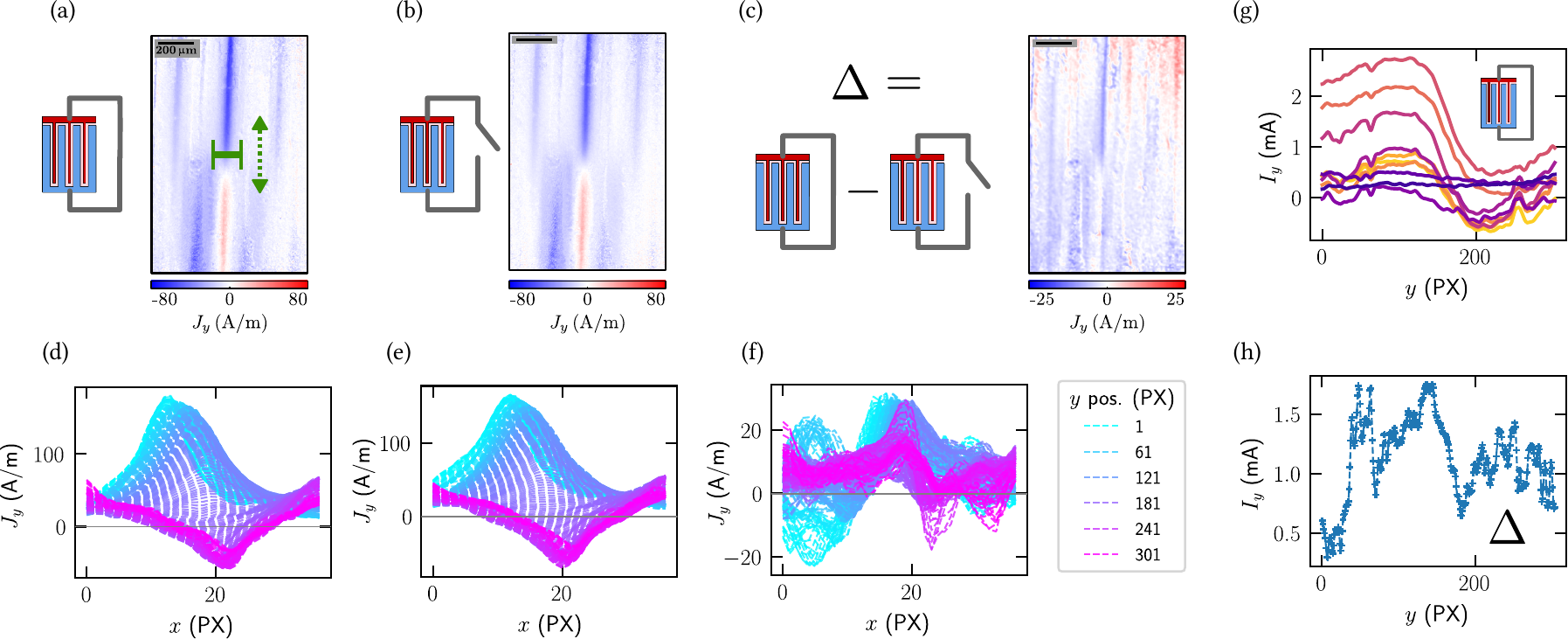}
	\caption{
		Isolating external current.
		Current density \(J_y\) images taken at end of IR pulse, under short-circuit (a) and open-circuit (b) conditions.
		(c)~Difference between short-circuit and open-circuit current densities, displaying the external path current density.
		(d-f)~Horizontal line cuts associated with (a-c).
		Each color is a profile taken for a fixed \(y\) position, for the width and horizontal location shown in (a).
        (g)~Each profile in (d) summed and plot against its \(y\) position, for each \(\tau\) value as in Fig.~\ref{fig:tseries}.
        (h)~Each profile in (e) summed and plot against its \(y\) position.
        Each image is 305\,PX\,\(\times\)\,201\,PX, at 4.33\,\(\upmu\)m/PX.}\label{fig:app_linecuts}
\end{figure*}

The short-circuit measurements, such as in Fig.~\ref{fig:tseries}, contain both purely internal current loops and external return currents. To isolate the external return currents, we compare the short-circuit measurement [Fig.~\ref{fig:app_linecuts}(a)] to an image under open-circuit conditions [Fig.~\ref{fig:app_linecuts}(b)].
The short-circuit and open-circuit currents have a very similar pattern, which is only evident when their subtraction is displayed [Fig.~\ref{fig:app_linecuts}(c)].
We analyze this system further through linecuts [Figs~\ref{fig:app_linecuts}(d-f)], where each series is the profile \(L_i(x) = J_y(x, y=i)\).
The short-circuit [Fig.~\ref{fig:app_linecuts}(d) and open-circuit [Fig.~\ref{fig:app_linecuts} profiles have the same shape except from a small horizontal offset.
Isolating this offset in Fig.~\ref{fig:app_linecuts}(f) produces a profile with less positive current, i.e.\ the upward current [red in Fig.~\ref{fig:app_linecuts}(a)] unexpected in the absence of internal current loops is diminished.
The short-circuit image line cuts are then integrated to produce the net vertical current as a function of \(y\) and time [Fig.~\ref{fig:app_linecuts}(g)].
By subtracting the open-circuit measurement [Fig.~\ref{fig:app_linecuts}(h)] we isolate a comparatively constant vertical current, identified with a relatively delocalized external current path.

\section{Stimulus from front of cell} \label{app:front}

\begin{figure*}[b!]
	\includegraphics[width=0.9\textwidth]{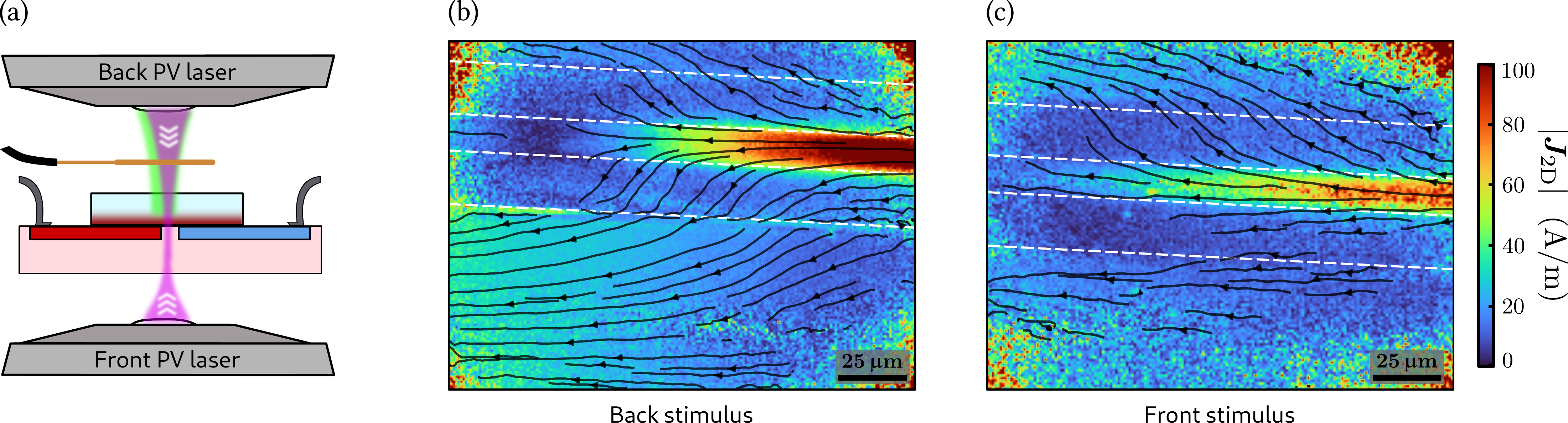}
	\caption{
		Comparison of stimulus from front and back of cell.
		(a)~Schematic of experiment, with PV laser directed either from above (back of cell) or below (front of cell) the device.
		(b,c)~Current density images of the IBC cell previously measured in Fig.~\ref{fig:tseries}, with stimulus from the back (b) and front (c) of the cell.}\label{fig:app_frontback}
\end{figure*}

We have reported photoresponse experiments with the optical stimulus oriented towards the back of each photovoltaic cell.
This decision allowed for increased current densities within the field of view, as well as a more two-dimensional current pattern.
However it may be beneficial or desirable in certain circumstances to image with stimulus from the standard side of the cell.
We tested this scenario by adjusting our microscope to optionally direct the PV laser correspondingly, with a second objective used to focus the beam [Fig.~\ref{fig:app_frontback}(a)].
Results taken under these two conditions are shown in Fig.~\ref{fig:app_frontback}(b) (back stimulus) and Fig.~\ref{fig:app_frontback}(c) (front stimulus).
We note that the front stimulus experiment has a more dispersed signal due to diffusion of the current as it passes vertically through the device.

\section{Imaging at maximum power point} \label{app:mpp}

\begin{figure}[tb!]
	\includegraphics[width=0.45\textwidth]{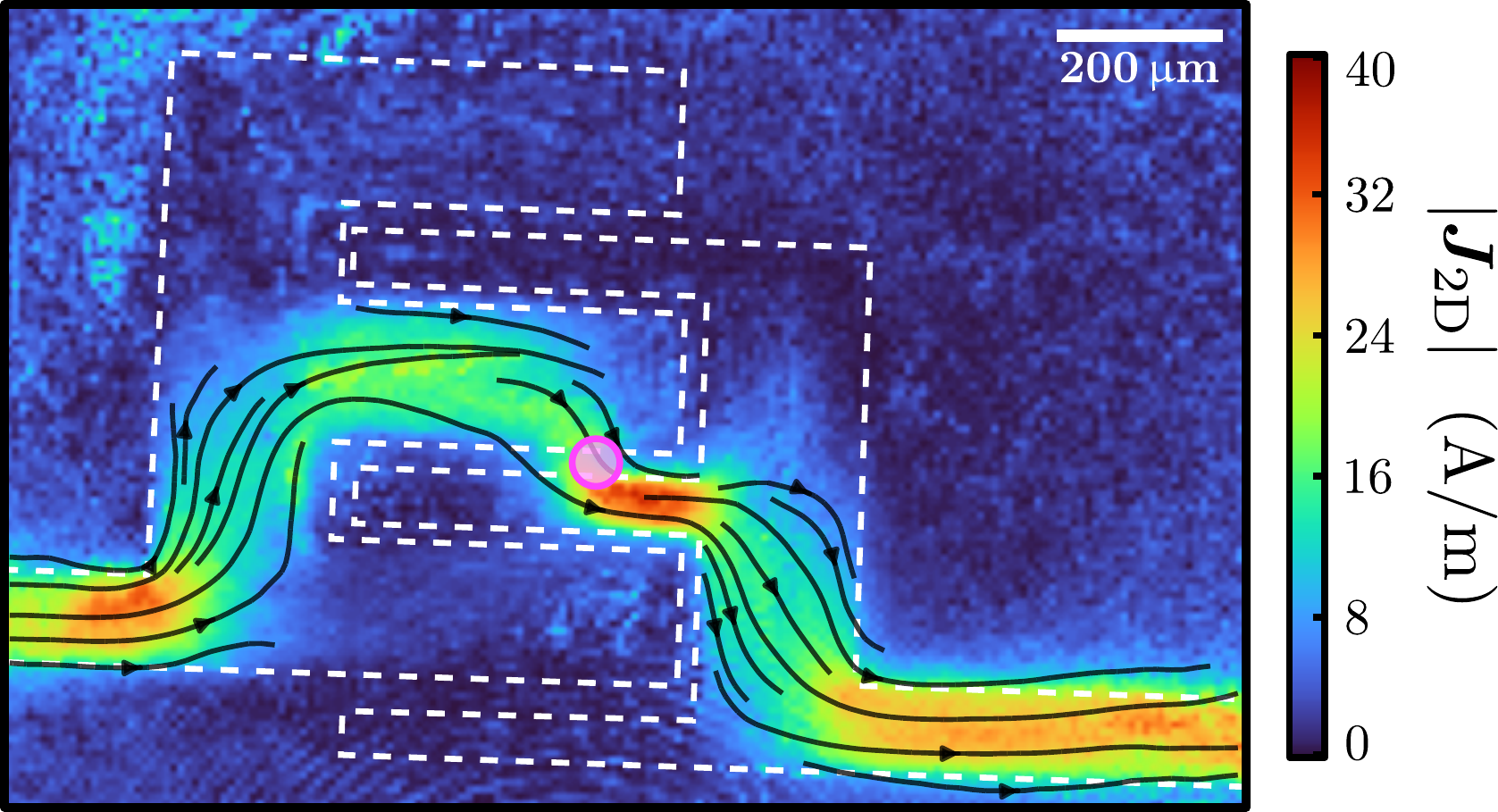}
	\caption{Current density image taken at maximum power point (MPP): 0.23\,V.}\label{fig:app_mpp}
\end{figure}

PV cell analysis techniques generally measure with the device under short or open circuit techniques where simpler operation can assist in  the extraction of useful information, particularly when comparing to theoretical models.
However neither of these points are where a PV cell operates at its highest power output, this `maximum power point' (MPP) occurs at the shoulder of the IV curve.
For the trident device studied in Fig.~\ref{fig:trident} this is at about 0.23\,V, and we show images taken at this potential [Fig.~\ref{fig:app_mpp}] to demonstrate this capability of the QDM platform.

\end{document}